\documentclass[nofootinbib,aps,prd,twocolumn]{revtex4-2}

\usepackage{amsmath}  
\usepackage{mathtools} 
\usepackage{amsfonts} 
\usepackage{graphicx} 
\usepackage{xcolor}
\usepackage{tikz}
\usepackage{pgfplots}
\pgfplotsset{compat=1.18}
\usetikzlibrary{arrows.meta}
\usepackage{amsthm}

\newcommand{\sminus}{\mathbin{\vcenter{\hbox{$\scriptstyle -$}}}}
\newcommand{\SchwSdS}{Schwarzschild-\mbox{(anti-)de Sitter }}

\begin{document}


\title{The equation of Binet in classical and relativistic orbital mechanics}

\author{Jose Luis Alvarez-Perez}
\email{joseluis.alvarez@uah.es}
\affiliation{Signal Theory and Communications Dept., Polytechnic
School, \ University of Alcala (Spain).}



\begin{abstract}
Binet's equation provides a direct way to obtain the geometric shape of orbits in a central force field. It is well known that in Newtonian gravitation Binet's equation leads to all the conic curves as solutions for an inverse-square force. Its standard derivation in higher-education textbooks relies on the use of polar coordinates and the conservation of angular momentum and energy. Here an alternative derivation is presented that emphasizes the physical interpretation of the problem by considering the motion as a superposition of a vertical fall toward the attracting centre plus an inertial motion in the horizontal and vertical directions. That is, we show how Binet's equation arises from the horizontal and vertical infinitesimal displacements of a body in free fall and in inertial motion. This deduction uses elementary concepts of infinitesimal calculus and connects closely with the original geometric and kinematical viewpoint of Newton and his predecessors and contemporaries. It is a view that is fully consistent with the maths taught in undergraduate courses. Second, we derive the relativistic version of Binet's equation for the \SchwSdS metric. This derivation, which is novel, directly relates the coordinates involved in Binet's equation without the need to introduce potentials or the use of Killing vectors. It is also an adequate step forward in the use of mathematical physics that can be introduced in advanced undergraduate or graduate courses of general relativity.
Finally, we tackle some controversies related to the role of the cosmological constant in the trajectory of photons in a \SchwSdS or even in Reissner--Nordström--(anti-)de Sitter spacetimes.  In this last step of the work, it is shown to students how a thorough yet concise mathematical analysis can solve disputes that arise even in high-level research articles. At the same time, the attentive reader is invited to connect the arguments provided at this final point with the analysis made in the case of non-relativistic Binet's equation.
\end{abstract}

\maketitle 

\section{Introduction} 

Among many others, two of Newton's foremost contributions were the
unification of the phenomena of the fall of heavy objects and the
motion of celestial bodies, on the one hand, and the systematization
of earlier infinitesimal methods into the foundations of modern
calculus, on the other. In regard to the motion of celestial
bodies, at the very start of his \textit{Principia} \cite{newton:1687}, Newton described the motion of the Moon as a superposition
of the motions due to inertia and gravitational fall. As for the infinitesimal
methods used in the \textit{Principia}, he used them under the idea of the so-called mathematical ultimate equality~\cite{needham:2021}. The modern use of these two leads naturally to the equations of motion in the two-body problem, as presented here.

Einstein developed a theory of gravitation that was based on the equivalence of the gravitational field and a uniformly accelerated reference frame. This principle of equivalence is intrinsically differential since it applies to an elemental neighbourhood of a point in spacetime where the gravitational field can be considered constant. Thus, we again have the two same ingredients in some manner: unification of different phenomena and differential calculus. In Einstein's case the latter consists in employing the tools of differential geometry, and understanding the motion of a body in a gravitational field as the motion along a geodesic line in a curved spacetime. 

The classical study of orbital motion in a two-body context under central forces is elegantly done with Binet's equation~\cite{gonzalez-lopez:2025}. This differential equation directly yields the geometric shape of the orbit without using time as a parametrization of the two-dimensional curves. It is named after Jacques Philippe Marie Binet (1786-1856), though it is often left unnamed in classical textbooks~\cite[p. 130]{symon:71}\cite[p. 87]{goldstein:2001}. 
For an inverse-square force, all conic curves are solutions of Binet's equation. The Newtonian Binet's equation has a counterpart in general relativity which includes additional terms.
Several approaches exist to derive it in either context, Newtonian \cite{symon:71,goldstein:2001} or Einsteinian \cite{misner:73,carroll:2019,schutz:2022}. Some original approaches have also been developed, such as in \cite{deliseo:2007}, which employs a complex-variable formalism. In the present article we obtain first the non-relativistic Binet's equation by making use of the two anticipated elements: inertia and the central force producing the dynamics, Newton's gravitational force in particular. The coordinate system is chosen to underline the idea of a vertical fall onto the attracting body and a given inertial speed. Traditionally, the problem is formulated from the outset in polar coordinates. Here, however, we begin by using a Cartesian coordinate system. Second, we derive the relativistic Binet's equation by directly relating the coordinates involved in it, that do not include time. This is different from the usual procedures and sheds light both on the tools of general relativity and its flexibility to adopt different coordinate systems.

\section{Binet's equation as a consequence of a vertical fall}
\label{sec-classical}
The formula known as the classical Binet's equation for the two-body problem and a central force field $f(r)$ \cite{symon:71,goldstein:2001,gonzalez-lopez:2025}
is given by
\begin{equation}
\frac{d^{2} u}{d \phi^{2}}+u = \frac{1}{\mu\, J^{2} u^{2}}f(1/u)
\label{eq-binet1}
\end{equation}
where $u=1/r$ is the inverse of the distance between the two mutually attracting bodies of masses $m_{1}$ and $m_{2}$, of which $\mu$ is the reduced mass, $J$ is the angular momentum of $\mu$ per unit mass and $\phi$ is the polar angle coordinate. The force $f(r)$ is, for the case of the Newtonian force of gravitational attraction,
\begin{equation}
f(r)= \frac{G\,m_{1}\,m_{2}}{r^{2}}
\end{equation}
where $G$ is the gravitational constant.

\begin{figure}
\centering
\begin{tikzpicture}[scale=1.2]
\def\offsetx{0.025}
\def\offsety{0.275}
\def\radius{1.125}
\def\a{2.75}
\def\b{1.75}
\def\c{sqrt(\a*\a-\b*\b)}
\def\anglerotellipse{0}
\coordinate (A) at (0,0.25);
\coordinate (B) at (-0.25,0.675);
\coordinate (C) at (-0.275,0.675);
\coordinate (D) at (-0.275,0.675);
\coordinate (P) at (0,1.125);
\coordinate (Pb) at (0++\offsetx,1.125+\offsety);
\draw[-Stealth] (0,0) -- ++(0,2) node[right] {\scalebox{1.1}{$y$}} ;
\draw[-Stealth] (0,0) -- ++(-2.75,0) node[left] {\scalebox{1.1}{$x$}} ;
\draw[-Stealth] (P) -- ++(0,0.4) node[xshift=10pt] {\scalebox{1.1}{$v_{y}$}} ;
\draw[-Stealth] (P) -- ++(-0.6,0) node[xshift=-2pt,yshift=8pt] {\scalebox{1.1}{$v_{x}$}} ;
\draw[very thin] (0,0) -- (-0.375,1.1);
\draw[very thin] (0,0.5) arc [start angle=90, end angle=109,
radius=0.5];
\node[below right] at (Pb) {$\mu$};
\node[right] at (A) {\scalebox{1.1}{$r$}};
\node[right] at (B) {\scalebox{0.9}{$\phi$}};
\draw[densely dotted] (0,0) circle [radius = \radius];
\draw[densely dotted, rotate={\anglerotellipse}] ({-\c*cos(\anglerotellipse)},{-\c*sin(\anglerotellipse)}) ellipse ({\a} and {\b});

\draw[densely dotted, domain=-3.75:1.25, samples=200]
      plot (\x, {1.125 - 0.2*\x*\x});
      
\filldraw [black] (0,0) circle (0.25pt);
\filldraw [black] (P) circle (1pt);
\end{tikzpicture}
\caption{Representation of the instantaneous position of the reduced mass $\mu$ in a Cartesian coordinate system $(x,y)$ centred at the attracting centre. The vertical $y$ axis is aligned with the direction from the attracting centre to $\mu$ at this instant. The horizontal $x$ axis is perpendicular to $y$. The initial velocity components $v_{x}$ and $v_{y}$ represent the inertial motion along $x$ and $y$, respectively. The dotted curves represent some possible orbits of $\mu$ around the attracting centre, depending on the initial conditions.}
\label{fig:orbits}
\end{figure}
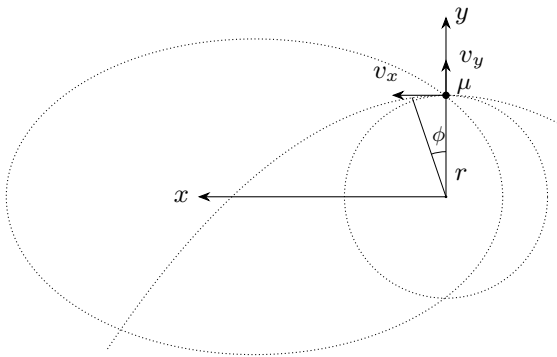

In Newtonian gravitation, the movement of a body in an orbit is the result of a combination of inertia and the gravitational pull by another massive object. Thus, for example, Newton explicitly describes the Moon as constantly falling toward the Earth, but always missing it because of its lateral velocity. Following this view, the derivation of \eqref{eq-binet1} presented here is based on four premises:
\begin{enumerate}
\item the equivalent one-body problem of a reduced mass placed in a central force field will be studied in a single, arbitrary instant and position of its trajectory; we will obtain Binet's equation from the local behaviour of the reduced mass at such instant and point,
\item the central nature of the force implies: i) that the motion of the reduced mass is confined to a plane defined by the force and the velocity vectors at the point and instant under consideration \footnote{This stems from the fact that there is no force component along the direction perpendicular to that plane.}, ii) that the description of the motion is independent of the axis chosen as the origin of the polar angle \footnote{At the point and instant of observation the initial conditions of position and velocity are the same for any choice of the origin of polar angles and the force does not depend on the polar angle.},
\item we will consider a reference frame defined by Cartesian coordinates, centred at the position of the attracting centre and oriented such that, at the instant of observation, the reduced mass is falling towards the centre of attraction along the vertical $y$ direction that connects both positions (see Figure \ref{fig:orbits}),
\item the inertia of the reduced-mass object is decomposed into vertical ($y$-axis–oriented) and horizontal ($x$-axis–oriented) components.  
\end{enumerate}
The choice of an inertial Cartesian system of coordinates $(x,y)$ stems from the fact that they naturally allow for the distinction between a vertical and a horizontal direction. This is the differential aspect of this demonstration in comparison to others \cite{templetracy:92,samboy:2024}. The normal approach is to write Newton's second law in polar coordinates from the very start due to the planar rotational symmetry of the problem \cite{samboy:2024}. Our approach starts from Cartesian coordinates to illustrate the physics of the vertical free fall but swiftly switches to polar coordinates.

The position of the reduced mass at $(x,y)$ is related to its polar coordinates by
\begin{align}
x&=r\,\sin\phi \nonumber
\\
y&=r\,\cos\phi
\end{align}

\noindent The infinitesimal displacement with time of this mass is given by $(dx,dy)$ and is represented in terms of the polar coordinates at the point under consideration by
\begin{align}
dx&=\sin\phi\,\dot{r}\,dt+r\,\cos\phi\,\dot{\phi}\,dt\nonumber
\\
dy&=\cos\phi\,\dot{r}\,dt-r\,\sin\phi\,\dot{\phi}\,dt
\label{eq-dxdy}
\end{align}
where the dot on top of a variable stands for the derivative with time. We now set $\phi=0$ as the direction of the $y$ coordinate, and the horizontal speed takes the form
\begin{equation}
\dot{x}=r\,\dot{\phi}
\end{equation}
This speed is related to the angular momentum per unit mass, $J$, by
\begin{equation}
\dot{x}=\frac{J}{r}
\label{eq-vxJr}
\end{equation}
and therefore we have
\begin{equation}
\dot{\phi}=\frac{J}{r^{2}}
\label{eq-thetaderivative}
\end{equation}

Regarding the vertical kinematics, we obtain
\begin{equation}
\ddot{y} =\ddot{r}\cos\phi-2\,\dot{r}\sin\phi\,\dot{\phi}-r\cos\phi\,\dot{\phi}^{2}-r\sin\phi\,\ddot{\phi}.
\end{equation}
With $\phi=0$, we have
\begin{equation}
\ddot{y}=\ddot{r}-r\,\dot{\phi}^{2}
\label{eq-ydoublederivative}
\end{equation}
If we consider a single rotation period, and consequently $\phi$ takes values between $0$ and $2\pi$, we can rewrite the corresponding derivatives as follows,
\begin{align}
\ddot{r}=-\frac{J^{2}}{r^{2}}\frac{d^{2}}{d\phi^{2}}\left(\frac{1}{r}\right)
\label{eq-rdoublederivative}
\end{align} 
Substituting equations \eqref{eq-thetaderivative} and 
\eqref{eq-rdoublederivative} into \eqref{eq-ydoublederivative}, replacing $r$ by its inverse, $u=1/r$, and introducing the dynamics of the central force $f(r)$ through the acceleration function $g(r)=f(r)/\mu$, we obtain
\begin{equation}
\frac{d^{2} u}{d \phi^{2}}+u =\frac{1}{J^{2} u^{2}} g(1/u)
\label{eq-binet2}
\end{equation}
which is equivalent to \eqref{eq-binet1}. For the case of Newtonian gravity, we have $g(r)=G M/r^{2}$, where $M= m_{1}+m_{2}$. Thus, \eqref{eq-binet2} becomes
\begin{equation}
\frac{d^{2} u}{d \phi^{2}}+u =\frac{GM}{J^{2}}
\label{eq-binet3}
\end{equation}
which is the Newtonian Binet's equation. Once solved for $u$ and then converted to $r$, it has all the conic curves as solutions:
\begin{equation}
r=\frac{p}{1+ e\,\cos(\phi-\phi_{0})}
\label{eq-conic-curves1}
\end{equation}
where $e$ is the eccentricity of the curve~\footnote{Values of $0 < e< 1$ correspond to an ellipse, $e=0$ to a circle, $e=1$ to a parabola and $e>1$ to a hyperbola.} and $p=J^{2}/GM$ is the semilatus rectum, i.e. the value of $r$ when $\phi-\phi_{0}=\pi/2$. As for $\phi_{0}$, it is the angle of the periapsis, and together with $e$, they form the set of integration constants determined by the initial conditions $(0,y)$ and $(v_{x},v_{y})$. In particular, we have
\begin{align}
u(\phi=0)&=\frac{GM}{y^{2}v_{x}^{2}}\left(1+e\,\cos\phi_{0}\right)\nonumber
\\
\left.\frac{d u}{d \phi}\right|_{\phi=0}&=\frac{GM\,e}{y^{2}v_{x}^{2}}\sin\phi_{0}
\label{eq-uandup}
\end{align}
which are related to the initial height and the horizontal and vertical inertia components through 
\begin{align}
u(\phi=0)&=\frac{1}{y}\nonumber
\\
\left.\frac{d u}{d \phi}\right|_{\phi=0}&=-\frac{v_{y}}{y\,v_{x}}.
\label{eq-uandup2}
\end{align}
where $v_{x}=\dot{x}=r\,\dot{\phi}$, $v_{y}=\dot{y}=\dot{r}$. The values of $y$, $v_{x}$ and $v_{y}$ are, by definition, initial values because the Cartesian coordinate system is considered and used here only at $t\in[0,dt]$. It is straightforward from \eqref{eq-uandup} and \eqref{eq-uandup2} to see that $y$, $v_{x}$ and $v_{y}$ determine $p$, $e$ and $\phi_{0}$~\footnote{Alternatively, the integration constants can be obtained from the angular momentum and the energy since $J=y\,v_{x}$ and $E=\frac{1}{2} (v_{x}^{2}+v_{y}^{2})-GM/y$}. The value of $GM$ determines the numerical mapping of the initial conditions onto these three parameters.

Mathematically, the most interesting features of the Binet's equation for the Newtonian attractive force are the following:
\begin{enumerate} 
    \item It is a linear ordinary differential equation (ODE), unlike the equation in $r=r(t)$ resulting from \eqref{eq-ydoublederivative} when introducing \eqref{eq-thetaderivative} and Newton's gravitational force,
    \begin{equation}
        \ddot{r}-\frac{J^{2}}{r^{3}}=-\frac{GM}{r^{2}}
    \label{eq-rdd}
    \end{equation}

    The term $J^{2}/r^{3}$ encodes the centrifugal force term in \eqref{eq-rdd}. The presence of a fictitious force in \eqref{eq-rdd} is due to the fact that the polar coordinate reference frame is a rotating reference frame~\footnote{One often finds recommendations on how to deal with fictitious forces that lack full rigour but exhibit a certain degree of practicality. For example, the text on \textit{Analytical Mechanics} by Hand and Finch~\cite{handfinch:98} recommends the following approach: ``Treat fictitious forces like real forces, and pretend that you are in an inertial frame''. Such prescriptions, however, are not conceptually optimal.} 
    

    \item The problem has been reduced to a one-dimensional problem with respect to the azimuthal angle, specifically the one of a free, undamped harmonic oscillator. The domain of $u=1/r$ is the positive real semi-axis, which makes it a half-line harmonic oscillator. If $0\le e < 1$, there is a periodic oscillation around $r=J^{2}/GM$, whereas for $e\ge 1$ there is a single pass trajectory. In the latter case, $u$ approaches 0, which corresponds to $r$ going to infinity. Similarly, any initial condition having $u(t_{0})>0$ always produces a positive solution for $r(\phi)$ because, to become negative, $u$ should traverse $0$ ($r=\infty$).
    \item It can be further reduced to 
    \begin{equation}
    \frac{d^{2} \tilde{u}}{d \phi^{2}}+ \tilde{u} = 0
    \label{eq-binet3b}
    \end{equation}
    after making the change of variables $\tilde{u}=u-GM/J^{2}$. The only parameter in equation~\eqref{eq-binet3b} is the angular frequency of the equivalent half-line simple harmonic oscillator, which is equal to one. This value of unity implies that the solution has a period of $2\pi$ in $\phi$. Otherwise, equation \eqref{eq-binet3b} does not contain any information about the inertia, nor does about its surrogate in \eqref{eq-rdd}, the centrifugal acceleration.

    If the initial conditions were $\tilde{u}_{0}=0$ and $\dot{\tilde{u}}_{0}=0$, the trajectory of the particle would be circular. Both $J$ and $GM$ have been absorbed into the change of variables. For non-circular orbits, the particle's position oscillates around $\tilde{u}_{0}=0$ in an ellipse or moves about it in a single pass in the cases of a parabola or a hyperbola.
    \end{enumerate}

In the next section, we will derive Binet's equation for the Schwarzschild-(anti-)de Sitter metric. Instead of a force-driven description, the motion is governed by the geodesic equations. A massive test particle follows a timelike geodesic whose four-velocity has norm $c$, whereas a photon follows a null geodesic whose tangent vector has zero norm. The geometry of the trajectory will not follow from the described combination of a horizontal and vertical momentum plus a vertical push. Instead, it will obey a geodesic path relating the first and second order derivatives of the position through the geometry of spacetime. This geometry is determined by a metric whose form is obtained as a solution of Einstein’s field equations. 

\section{Binet's equation with Schwarzschild-(anti) de Sitter metric}
As is well known, the physical meaning of the motion of an object in a gravitational field changes fundamentally in general relativity compared to Newton's theory. Two of the most important differences are the shift from the concept of force to that of spacetime curvature, and the nonlinearity implied by Einstein’s field equations. The latter means that, unlike in Newtonian gravitation, the two-body problem does not reduce to a one-body problem because the necessary superposition principle does not hold. However, if one of the masses is much larger than the other, it is possible to treat it as a one-body problem where the smaller mass moves in the gravitational field created by the larger, stationary mass. This is the scenario we consider here. Regarding the first point, the concepts of force and inertia are replaced by the motion of the object along a geodesic in curved spacetime. Such a geodesic is determined by the metric that solves Einstein's equations for the gravitational field created by the massive body.

Our approach in section \ref{sec-classical} was to portray the movement of a body in the presence of an attractive field as a free fall that is missed due to its inertia, resulting in an orbit. In the context of general relativity, the motion known as \textit{free fall} is not produced by the gravitational force attracting the massive particle but by the modification of spacetime curvature, so that the aforementioned geodesic must be followed. A geodesic is the trajectory of an object in spacetime when no other forces are present~\footnote{The expression 'free fall' in classical and relativistic mechanics refers to the same phenomenon but with important nuances. In both cases, it refers to the absence of non-gravitational forces; however, in the classical case, free fall occurs toward the centre of mass, whereas in general relativity, the gravitational picture is replaced by curved spacetime and such a free fall is a slide along the geodesic with a constant four-velocity magnitude.}. Regarding gravitation, the body is not 'forced' to follow the $y$ direction toward the attractive centre as in the analysis presented in section \ref{sec-classical}. Instead, the infinitesimal displacement now occurs such that the covariant change $D$ of the tangent four-vector along the geodesic, $v^{\mu}$, is zero, as defined by
\begin{equation}
D v^{\mu} \equiv dv^{\mu}+\Gamma_{\nu\sigma}^{\mu} v^{\sigma} dx^{\nu}=0
\label{eq-geodesicseq1}
\end{equation}
with
\begin{equation}
v^{\mu}\equiv\frac{dx^{\mu}}{d\ell}
\end{equation}
where $x^{\mu}$ is the spacetime four-position, $(c\,t,x,y,z)$, $\Gamma_{\nu\sigma}^{\mu}$ are the Christoffel symbols and $\ell$ is the length element of the geodesic in spacetime, to be defined more precisely below. The Christoffel symbols describe how the coordinate basis changes from point to point in curved spacetime. Both sides of equation \eqref{eq-geodesicseq1} can be divided by $d\ell$ to produce
\begin{equation}
\frac{d v^{\mu}}{d\ell} +\Gamma_{\nu\sigma}^{\mu} v^{\sigma} v^{\nu}=0.
\label{eq-geodesicseq1b}
\end{equation}
\noindent The length of an element of a curve such as a geodesic in the Lorentzian space of signature $(-+++)$ is the square root of the absolute value of the quadratic form $ds^{2}$,
\begin{equation}
d\ell = \sqrt{|ds^{2}|}.
\end{equation}
The metric $ds^{2}$ is defined as
\begin{equation}
ds^{2}=g_{\mu\nu}\, dx^{\mu} dx^{\nu}
\end{equation}
where $g_{\mu\nu}$ is the metric tensor. For a subluminal particle the quadratic form $ds^{2}$ is negative. Thus, in the particle's rest frame we have $ds^{2}=-c^{2} d\tau^{2}$ where $\tau$ is the proper time and $d\ell=c\,d\tau$. For a photon, $ds^{2}=0$ and hence $d\ell=0$. In the latter case a different parametrization of the geodesic is required, which must still satisfy \eqref{eq-geodesicseq1b}, a fact that makes such a parametrization \textit{affine} by definition. The proper time $\tau$ is an example of an affine parameter, in the case of a massive particle. Thus, the general geodesic equation can be written as
\begin{equation}
\frac{d^{2} x^{\mu}}{d\lambda^{2}} +\Gamma_{\nu\sigma}^{\mu} \frac{d x^{\sigma}}{d\lambda} \frac{d x^{\nu}}{d\lambda}=0.
\label{eq-geodesicseq1c}
\end{equation}
for any such affine parameter~\footnote{Equation \eqref{eq-geodesicseq1c} is invariant under any affine transformation $\lambda=a S+b$ with $a,b\in\mathbb{R};a\neq 0$, which gives it its name.}. Equation \eqref{eq-geodesicseq1c} 
replaces Newton's second law for a massive particle and free-fall motion when $\lambda=\tau$.

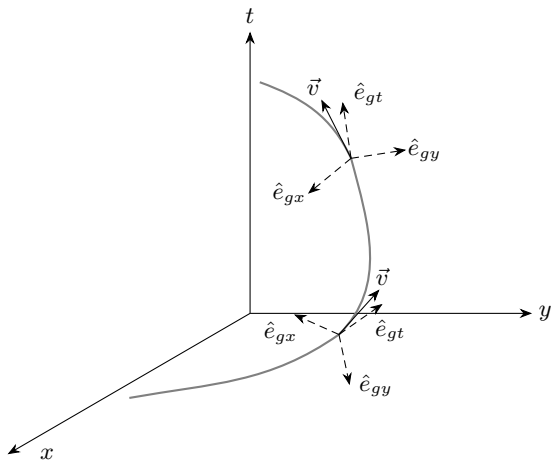
\begin{figure}
\centering
\begin{tikzpicture}[x={(0.86cm,0.5cm)},{y={(1cm,0cm)}},z={(0cm,1cm)},
scale=1.5,
    hypersurf/.style={fill=green!10,draw=green!50!black,thick},
    triad/.style={->,thick,blue!60!black}]
\def\offsety{0.5}
\def\slantxAB{0.9}
\def\slantyAB{-0.9}
\def\slantxB{-0.15}
\def\slantyB{0.15}
\coordinate (A) at (0.5,-2+\offsety,-1);
\coordinate (B) at (0.75,-0.25+\offsety,1);
\coordinate (C) at (0.25,0+\offsety,1.5);
\coordinate (D) at (0.1,-0.5+\offsety,2.);
\coordinate (AB) at (0.625,-0.25+\offsety,-0.5);

\draw[-Stealth] (0,0,0) -- (-2.5,0,0) node [xshift=15pt] {$x$};
\draw[-Stealth] (0,0,0) -- (0,2.5,0) node [xshift=5pt] {$y$};
\draw[-Stealth] (0,0,0) -- (0,0,2.5) node [above] {$t$};

\draw[gray,thick] 
  (A) 
  to[out=10,in=-145] (AB)
  to[out=45,in=-75] (B)
  to[out=115,in=-20] (D)
  ;

\draw[-Stealth,densely dashed,xslant=\slantxAB,yslant=\slantyAB] (AB) -- ++(-0.65,0,0) node [xshift=-5pt,yshift=-7pt] {$\hat{e}_{gx}$};
\draw[-Stealth,densely dashed,xslant=\slantxAB,yslant=\slantyAB] (AB) -- ++(0,0.5,0) node [xshift=10pt] {$\hat{e}_{gy}$};
\draw[-Stealth,xslant=\slantxAB,yslant=\slantyAB] (AB) -- ++(0,0,0.4) node [xshift=1pt,yshift=5pt] {$\vec{v}$};
\draw[-Stealth,densely dashed,xslant=\slantxAB,yslant=\slantyAB] (AB) -- ++(0.05,0.1,0.375) node [xshift=3pt,yshift=-10pt] {$\hat{e}_{gt}$};

\draw[-Stealth,densely dashed,xslant=\slantxB,yslant=\slantyB] (B) -- ++(-0.5,0,0) node [xshift=-7pt] {$\hat{e}_{gx}$};
\draw[-Stealth,densely dashed,xslant=\slantxB,yslant=\slantyB] (B) -- ++(0,0.5,0) node [xshift=7pt] {$\hat{e}_{gy}$};
\draw[-Stealth,densely dashed,xslant=\slantxB,yslant=\slantyB] (B) -- ++(0,0,0.5) node [xshift=10pt,yshift=3pt] {$\hat{e}_{gt}$};
\draw[-Stealth,xslant=\slantxB,yslant=\slantyB] (B) -- ++(-0.1,-0.1,0.6) node [xshift=-3pt,yshift=5pt] {$\vec{v}$};
  
\end{tikzpicture}
\caption{A massive particle follows a trajectory (solid curve) in curved spacetime determined by the metric $g_{\mu\nu}$. At each point of the trajectory, a local Lorentz frame (dashed arrows) can be constructed, in which the metric is locally flat. The tangent vector to the trajectory, $\vec{v}$, is parallel transported along the curve, meaning that its covariant derivative is zero. This parallel transport ensures that the particle's motion is consistent with the curvature of spacetime.}
\label{fig:geodesic}
\end{figure}

Figure \ref{fig:geodesic} illustrates the concept of a geodesic in curved spacetime. At each point of the spacetime -and therefore at each point along the geodesic- a so-called \textit{local Lorentz frame}, i.e. a locally flat tangent space, can be constructed. One way to understand the geodesic equation is to note that it expresses the condition under which the tangent vector remains constant -or, better, \textit{parallel transported}- along the curve. This parallel-transport condition must hold true in free fall for all points on the geodesic, as viewed from their respective local Lorentz frames.

To derive the relativistic version of Binet's equation, it seems natural to modify the geodesic equation for non-affine parameters \cite[p.~109]{carroll:2019}, in particular taking the azimuthal angle $\phi$ as parameter. The aim is to express the geodesic equation in terms of the inverse radial coordinate $u=1/r$ and the angular coordinate $\phi$. This approach is an alternative to the usual procedure of writing the geodesic equation in terms of the radial coordinate and the proper time, and then substituting $r$ with $u$ and $\tau$ with $\phi$ \cite{misner:73,schutz:2022,carroll:2019}. This would mirror the path used in the derivation of Binet's equation in the non-relativistic case. The direct use of $u$ and $\phi$ is possible because the geodesic equation can accommodate a non‑affine parameter. As shown in Appendix A, the geodesic equation with $\phi$ as parameter is 
\begin{equation}
\frac{d^{2} x^{\mu}}{d\phi^{2}}+\Gamma_{\nu\sigma}^{\mu} \frac{d x^{\sigma}}{d\phi} \frac{d x^{\nu}}{d\phi}=-\kappa(\phi) \frac{d x^{\mu}}{d\phi}.
\label{eq-geodesicseq1c2}
\end{equation}
The function $\kappa(\phi)$ is defined in Appendix A and is related to the first and second order derivatives of $\phi$ with respect to the affine parameter, in particular to the proper time for a timelike geodesic. Hence a hidden dependence on the proper time remains when using a non-affine parameter. Consequently, we will combine the affine and non-affine forms of the geodesic equation. Before proceeding further, we must specify the spacetime geometry in which the geodesic is traced.

In this work we choose the Schwarzschild-(anti-)de Sitter metric\footnote{By this we mean both the Schwarzschild--de Sitter and Schwarzschild--anti--de Sitter metrics, corresponding to positive and negative cosmological constants, respectively.}, which is the simplest spacetime geometry of the vacuum outside a static, spherically symmetric massive body with a nonzero cosmological constant. It is therefore a solution to Einstein equations with $\Lambda\neq 0$,
\begin{equation}
R_{\mu\nu}-\frac{1}{2} g_{\mu\nu}\,R+\Lambda g_{\mu\nu}=\frac{8\pi G}{c^{4}} T_{\mu\nu}
\end{equation}
where $R_{\mu\nu}$ is the Ricci curvature tensor, $R$ the Ricci scalar, $g_{\mu\nu}$ the metric tensor, $T_{\mu\nu}$ the stress-energy tensor, $G$ the gravitational constant, and $c$ the speed of light in vacuum~\footnote{We use physical units in this article, even though it is more common to choose geometrized units in which $G=c=1$ so that they do not appear in the equations. The latter option yields masses measured in length units. To recover physical units one replaces $M$ by $G M/c^{2}$}. In the vacuum outside the massive body, we have $T_{\mu\nu}=0$ and the solution is given by the Schwarzschild-(anti-)de Sitter metric, namely~\footnote{The Schwarzschild–(anti)–de Sitter solution is also called Kottler solution.}
\begin{equation}
ds^{2}=-f(u)c^{2} dt^{2}+\frac{1}{u^{4} f(u)} du^{2}+\frac{1}{u^{2}} d\theta^{2}+\frac{1}{u^{2}}\sin^{2}\theta\, d\phi^{2}
\label{eq-schwartzschild1a}
\end{equation}
with
\begin{align}
f(u&)=1-r_{S}\,u-\frac{\Lambda}{3\,u^{2}}\nonumber
\\
r_{S}&=\frac{2\, GM}{c^{2}}
\label{eq-schwartzschild1b}
\end{align}
where $r_{S}$ is the Schwarzschild radius. The function $f(u)$ is not simply an abbreviation for the coefficient of $c^{2} dt^{2}$ in the metric. Noting that $d\tau^{2}=-1/c^{2}\,ds^{2}$ defines the proper time, we see that $f(u)$ determines the relation between this proper time and the coordinate time increment $dt$ at a fixed spatial point, $d\tau=\sqrt{f(u)}\,dt$, i.e. when $du$, $d\theta$, and $d\phi$ are all zero~\footnote{This is the well-known effect that a clock runs more slowly at smaller radial distances from a gravitational mass.}. After computing the Christoffel symbols for this metric (see Appendix C) and setting $\theta=\pi/2$ (equatorial orbit plane) due to the symmetry of the problem, the non-affine geodesic equations take the form:
\begin{subequations}
\label{eq-geodesic-phi}
\begin{align}
\frac{d^{2}t}{d\phi^{2}}+\frac{f'(u)}{f(u)}\frac{d t}{d\phi}\frac{d u}{d\phi}&=\sminus\kappa(\phi) \frac{d t}{d\phi}
\label{eq-geodesic-t-phi}
\\
\frac{d^{2}u}{d\phi^{2}}+\frac{1}{2}c^{2} u^{4}\,f(u)\,f'(u)\left(\frac{d t}{d\phi}\right)^{2}&
\\
\sminus\!\left[\frac{1}{2} \frac{f'(u)}{f(u)}+\frac{2}{u}\right]\left(\frac{d u}{d\phi}\right)^{2}\!+
u f(u)&=\sminus\kappa(\phi)\frac{d u}{d\phi} \nonumber
\\
\sminus\frac{2}{u}\frac{d u}{d\phi}&=\sminus\kappa(\phi)
\label{eq-geodesic-u-phi}
\end{align}
\end{subequations}
where $\kappa(\phi)$ can be eliminated by substituting the last equation into the others to yield
\begin{subequations}
\label{eq-geodesic-phi2}
\begin{align}
\frac{d^{2}t}{d\phi^{2}}&+\left[\frac{2}{u}+\frac{f'(u)}{f(u)}\right] \frac{d t}{d\phi}\frac{d u}{d\phi}=0
\label{eq-geodesic-t-phi2}
\\
\frac{d^{2}u}{d\phi^{2}}+\frac{1}{2}c^{2} u^{4} &\,f(u)\,f'(u)\left(\frac{d t}{d\phi}\right)^{2} \nonumber
\\
\sminus & \frac{1}{2} \frac{f'(u)}{f(u)} \left(\frac{d u}{d\phi}\right)^{2}\!+
u f(u)=0
\label{eq-geodesic-u-phi2}
\end{align}
\end{subequations}
The dependence on the proper time is not present in equations \eqref{eq-geodesic-phi2} as it was in \eqref{eq-geodesic-phi} through $\kappa(\phi)$. The result has been the reduction of the system to two equations. Apparently, equations \eqref{eq-geodesic-phi2} may seem to be the geodesic equations in a different, corrected metric with two coordinates, $t$ and $u$, where $\phi$ would play the role of the affine parameter and where one dimension had been removed. This would be an outstanding result but the reasoning would be incomplete and flawed: we still have the pseudo-normalization condition described in Appendix B. The normalization condition for the four-velocity in the case of an affine parametrization or its pseudo-version for the non-affine case is not an extra condition but rather a necessary consequence of the geodesic equations (see Appendix B). However, having lost one of the geodesic equations, such (pseudo-)normalization condition is now not guaranteed by the remaining equations and must be imposed -or kept- separately.

The procedure is now to use \eqref{eq-geodesic-t-phi2} and integrate it as explained in Appendix D to produce \eqref{eq-conservation-t-phi}, which allows $dt/d\phi$ to be replaced in \eqref{eq-geodesic-u-phi2}. In addition to this, the square of $d u/d\phi$ in the latter can be calculated from the pseudo-normalization condition \eqref{eq-norm-wmu}, with $\alpha=\phi$ and $\lambda=\tau$ (or simply $\lambda$ for photons), and again from \eqref{eq-conservation-t-phi} as in Appendix D, producing equation \eqref{eq-du-dphi-squared}. After all these substitutions, equation \eqref{eq-geodesic-u-phi2} becomes~\footnote{The involved expressions in Appendices D and E contained an integration constant $B$ that is not in \eqref{eq-binet-relativistic0} as it gets canceled along the way.}
\begin{equation}
\frac{d^{2} u}{d\phi^{2}}+f(u)\,u=\begin{cases} \sminus\dfrac{1}{2}f'(u)\, u^{2}\left[1+c^{2}u^{2}\left(\dfrac{d\tau}{d\phi}\right)^{2}\right] & \mbox{(particle)} \\ \\ \sminus\dfrac{1}{2}f'(u)\, u^{2} & \mbox{(photon)} \end{cases}
\label{eq-binet-relativistic0}
\end{equation}
These equations may be referred to as 'pre-Binet'. They are not specific to the Schwarzschild--(anti-)de Sitter metric, since the function $f(u)$ has not yet been fixed as in \eqref{eq-schwartzschild1b}. Appropriate choices of $f(u)$ reproduce the Reissner--Nordström and Reissner--Nordström--(anti-)de Sitter spacetimes, as well as numerous static, spherically symmetric solutions that appear in modified gravity theories (for example, Einstein--Born--Infeld or Gauss--Bonnet models).

If we now substitute $f(u)$ into the \SchwSdS metric, as well as inserting \eqref{eq-conservation-angularmomentum-tau}~\footnote{The only input from the affine geodesic equations comes from the conservation law given by \eqref{eq-conservation-angularmomentum-tau}. It effectively compensates for the disappearance of equation \eqref{eq-geodesic-u-phi}.}, we obtain
\begin{equation}
\frac{d^{2} u}{d\phi^{2}} + u = \begin{cases} \dfrac{r_{S}\,c^{2}}{2\,J^{2}}+\dfrac{3}{2} r_{S} u^{2} -\dfrac{\Lambda c^{2}}{3 J^{2} u^{3}} & \mbox{(particle)} \\ \\ \dfrac{3}{2} r_{S} u^{2} & \mbox{(photon)} \end{cases} 
\label{eq-binet-relativistic1}
\end{equation}

These relativistic Binet's equations are usually derived in textbooks by differentiating equation \eqref{eq:first-int} with respect to $\phi$. In fact, as explained in Appendix F, equation \eqref{eq:first-int} is of a type for which a general analytical solution is known, unlike equation \eqref{eq-binet-relativistic1}. This seems to put Binet's equation behind \eqref{eq-du-dphi-squared} in importance. However, it is still relevant to have it at hand. First, it is interesting to confront the relativistic version of Binet's equation with the non-relativistic one. Thus, we note the presence of the relativistic correction terms proportional to $u^{2}$ and $u^{-3}$. These terms are responsible for the non-closure of elliptical orbits and the consequent precession of the perihelion in an elliptical orbit in general relativity. This is consistent with Bertrand's theorem~\cite{goldstein:2001}.  This theorem states that Newton's inverse-square and Hooke's laws are the only central forces in equation \eqref{eq-binet1} for which all bounded orbits are closed. To obtain a quantitative estimate of this effect, we could consider the case of low-eccentricity elliptical orbits. It is straightforward to show that a linearization of the nonlinear terms leads to a harmonic oscillator whose angular period differs from $2\pi$, implying that the orbit is not closed. In effect, by assuming that the ellipse is a deformation of a circle of inverse radial coordinate $u_{0}$, we can write
\begin{equation}
\frac{d^{2} u}{d\phi^{2}} + \mathcal{A} u = \mathcal{B}
\label{eq-binet-relativistic-approx}
\end{equation}
with
\begin{align}
\mathcal{A} &\approx \begin{cases} 1 - 3 r_{S} u_{0} - \dfrac{\Lambda\, c^{2}}{J^{2}\, u_{0}^{4}} & \mbox{(particle)} \\ \\
1 - 3 r_{S} u_{0} & \mbox{(photon)}\end{cases} \nonumber
\\
\\
\mathcal{B} &\approx \begin{cases} \dfrac{r_{S} c^{2}}{2 J^{2}} - \dfrac{3}{2} r_{S} u_{0}^{2} - \dfrac{4\,\Lambda\, c^{2}}{3\, J^{2}\, u_{0}^{3}} & \mbox{(particle)} \\ \\ 
\sminus\frac{3}{2} r_{S} u_{0}^{2} & \mbox{(photon)}\end{cases} \nonumber
\end{align}
As anticipated, the deviation of $\mathcal{A}$ from unity gives rise to the precession of the perihelion.

Some controversy arose in the literature when Binet's equation for photons in the Schwarzschild-(anti-)de Sitter metric was analyzed by Islam in 1983 \cite{islam:83}. He claimed that, since the cosmological constant $\Lambda$ did not appear in \eqref{eq-binet-relativistic1}, it had no effect on the trajectory of photons. This claim was refuted by Rindler and Ishak in 2007 \cite{rindler:2007}, who defined some observables that did depend on $\Lambda$. Further discussion was conducted by Arakida and Kasai in 2012 \cite{arakida:2012} on the basis that the absence of  $\Lambda$ in the geodesic equation of light was a misleading statement. They emphasized the role of equation \eqref{eq:first-int} in showing the actual dependence of the trajectory on $\Lambda$. As explained in Appendix F, the solution of \eqref{eq-binet-relativistic1} is given by the classic formula \cite[p.~87, eq.~(54)]{hagihara:31}~\footnote{Although the exact integration of Schwarzschild geodesics in terms of Weierstrass elliptic functions is credited to Hagihara \cite{hagihara:31}, the essential mathematical structure was already present in Whittaker's \textit{Analytical Dynamics} \cite{whittaker:17}. In this work, Whittaker classified central-force motions by the polynomial degree of the effective radial equation and showed that the general solutions of the cubic and quartic cases are expressible through Weierstrass elliptic functions. Schwarzschild orbits fit precisely into this class, so Whittaker's analysis anticipates the later relativistic development. Whittaker's work, however, is not cited by Hagihara, who thus appears to have rediscovered the method independently.}
\begin{equation}
r(\phi)=\frac{r_{S}}{4\, \wp\left(\phi-\phi_{in}\right)+\frac{1}{3}} 
\label{eq-solrelativisticlight}
\end{equation}
where $\wp(z)$ is the Weierstrass elliptic function and $\phi_{in}$ is an azimuthal offset for this function. This offset is such that the initial conditions are satisfied (in a similar way to equations \eqref{eq-conic-curves1}, \eqref{eq-uandup} and \eqref{eq-uandup2} for essentially the same mathematical problem). The solution for the massive particle is significantly more involved and can be found in \cite[p.~7, eq.~(63)]{hackmann:2008}. As for the aforementioned controversy, equation \eqref{eq-solrelativisticlight}, which is the solution of both \eqref{eq-binet-relativistic1} and \eqref{eq-du-dphi-squared} in the lightlike case, depends on $\Lambda$ (see Appendix F). Equation \eqref{eq-binet-relativistic1} has other hidden dependencies apart from $\Lambda$, such as the one upon the integration constant $B$ and therefore upon the energy and the angular momentum. The initial conditions map onto the solutions of the equations of motion, or geodesic equations in this case, through these apparently hidden constants such as $\Lambda$. It is also instructive in this regard to examine \eqref{eq-binet3b} in the non-relativistic case and observe the absence of the parameters $GM$ and $J$ in it. Nevertheless, the solution of \eqref{eq-binet3b} depends on them through the definition of the measured quantity, $\tilde{u}$, and through the manner in which the initial conditions determine the eccentricity, $e$, the semilatus rectum, $p$, and the angle of the periapsis, $\phi_{0}$.

\section{Conclusions}
In Newtonian mechanics, inertial frames are usually preferred because Newton’s laws hold in them without the introduction of fictitious forces. In general relativity, affine parametrizations are recommended for rather different reasons. However, both choices share the aim of avoiding additional and somewhat non-physical terms in the equations of motion. Nonetheless, we have ended up using a non-inertial frame -the one of polar coordinates- in Newtonian mechanics and non-affine parametrizations in general relativity to derive Binet's equation. This procedure is advantageous in the case of the non-relativistic Binet's equation because it reproduces the historical derivation and the original physical intuition behind it. It begins from a differential analysis of the trajectory in Cartesian coordinates and converts it to polar coordinates, which constitute the most convenient system for central-force problems. In the relativistic case, the use of the inverse radial coordinate and the azimuthal coordinate as a non-affine parameter from the outset allows for a direct derivation of Binet's equation in terms of the coordinates that were found to be ideal in the Newtonian case.

Binet's equation in its most general form is a one-degree-of-freedom autonomous conservative ODE~\footnote{An ODE described by $d^{2}\xi/d\alpha^{2}=\sum_{n=-M}^{N} \xi^{n}\; ;M,N\in\mathbb{N}$ is autonomous for not including the dependent variable $\alpha$ and conservative because of the integrability of the Laurent polynomial.}, both in the non-relativistic and relativistic cases. 
This property makes it particularly suitable for various types of analysis of the motion. We have offered a complementary perspective for the relativistic case in which conservation quantities such as the energy and angular momentum arise naturally from the geodesic equations with no need to invoke potentials or Killing vectors. 

The approach followed here is indeed adequate for introducing Binet's equation in a general relativity course after the geodesic equations have been derived. It is also a useful exercise that allows us to address controversies such as the one related to the influence of the cosmological constant on the trajectory of photons.

\appendix 
\section{Geodesic equation for a non-affine parameter}
\label{sec-nonaffine}

We start from the covariant geodesic law written for an affine parameter $\lambda$, as in equation \eqref{eq-geodesicseq1c}. For any function $\alpha=\alpha(\lambda)$, the chain rule gives
\begin{align}
\frac{dx^\mu}{d\lambda} &= \frac{dx^\mu}{d\alpha}\frac{d\alpha}{d\lambda},\\
\frac{d^2x^\mu}{d\lambda^2} &= \frac{d^2x^\mu}{d\alpha^2}\!\left(\frac{d\alpha}{d\lambda}\right)^{\!2}
+\frac{dx^\mu}{d\alpha}\frac{d^2\alpha}{d\lambda^2}.
\end{align}
Substituting these into the affine geodesic equation and dividing by $(d\alpha/d\lambda)^2$ yields the non-affine form
\begin{equation}
\frac{d^2 x^\mu}{d\alpha^2}+\Gamma^{\mu}_{\nu\sigma}\frac{dx^\sigma}{d\alpha}\frac{dx^\nu}{d\alpha}
= -\kappa(\alpha)\,\frac{dx^\mu}{d\alpha},
\label{eq-nonaffine-geodesic}
\end{equation}
where the \textit{non-affinity function} is
\begin{equation}
\kappa(\alpha)=\sminus\frac{d}{d\alpha}\left(\ln\frac{d\lambda}{d\alpha}\right)
\label{eq-kappadef1}
\end{equation}
or also
\begin{equation}
\kappa(\alpha)=\frac{d^2 \alpha}{d\lambda^2}\left(\frac{d\alpha}{d\lambda}\right)^{-2}.
\label{eq-kappadef2}
\end{equation}

\section{Norm of the tangent vector of a geodesic}
\label{sec-normalizationcondition}
The norm of the tangent vector is constant along a geodesic if defined with respect to an affine parametrization given by $\lambda$ as in equation \eqref{eq-geodesicseq1c}. This conservation principle is a consequence of the geodesic equation itself, which we rewrite here with a more concise notation, which follows the one in \cite[p.~105]{carroll:2019}:
\begin{align}
\frac{D\, v^{\mu}}{d\lambda}&=0
\label{eq-geodesic-affine-covariant-derivative}
\\
v^{\mu}&\equiv \frac{d x^{\mu}}{d\lambda}\nonumber
\end{align}
with $D/d\lambda$ being the covariant derivative along the curve. Thus, by applying this operator to the tangent vector, we have~\footnote{The covariant derivative of a scalar is equal to its ordinary derivative and, more importantly, it is devised to obey the Leibniz rule independently of the nature of the objects it is applied to.}
\begin{equation}
\frac{D}{d\lambda}\left(v^{\mu} v_{\mu}\right)= 2 v_{\mu} \frac{D v^{\mu}}{d\lambda}=0
\end{equation}
As a result of this, the squared norm of the tangent vector $v^{\mu}$ is constant along the geodesic. This constant equals $-c^{2}$ for massive particles and $0$ for photons, reflecting the definition of proper time and the fact that photons lie on the boundary of spacetime's causal structure. 

As shown in Appendix A, the geodesic equation for a non-affine parameter $\alpha$ is given by
\begin{equation}
\frac{D\, w^{\mu}}{d\alpha}=-\kappa(\alpha)\, w^{\mu}
\end{equation}
with
\begin{equation}
w^{\mu}\equiv \frac{d x^{\mu}}{d\alpha}.
\end{equation}
By applying the covariant derivative on the squared norm of $w^{\mu}$, we obtain
\begin{equation}
\frac{D}{d\alpha}\left(w^{\mu} w_{\mu}\right)= 2 w_{\mu} \frac{D w^{\mu}}{d\alpha}=-2 \kappa(\alpha) w_{\mu} w^{\mu}
\end{equation}
which shows that the squared norm of $w^{\mu}$ is not conserved along the geodesic unless $\kappa(\alpha)=0$, i.e., unless $\alpha$ is an affine parameter. Considered as a differential equation for the squared norm of $w^{\mu}$, its solution is
\begin{equation}
w^{\mu} w_{\mu}=C\, e^{-2 \int \kappa(\alpha) d\alpha}=C\,\left(\frac{d\lambda}{d\alpha}\right)^{2}
\label{eq-norm-wmu}
\end{equation}
where $C$ is a constant and where \eqref{eq-kappadef1} has been used. For the case $\lambda=\tau$ (or simply $\lambda$ for photons) and $\alpha=\phi$, we have $C=-c^{2}$ (massive particles) or $C=0$ (photons). The pseudo-normalization condition \eqref{eq-norm-wmu} could have been anticipated from $w^{\mu}=v^{\mu}\left(d\lambda/d\phi\right)$ and the analogous expression for $w_{\mu}$. However, the intention here was to show how it arises from the non-affine geodesic equation.

\section{Derivation of the Christoffel symbols of the Schwarzschild-(anti-)de Sitter metric}
From the metric matrix
\begin{equation}
g_{\mu\nu}=\begin{pmatrix}
-f(u)\,c^{2} & 0 & 0 & 0 \\
0 & \dfrac{1}{u^{4} f(u)} & 0 & 0 \\
0 & 0 & \dfrac{1}{u^{2}} & 0 \\
0 & 0 & 0 & \dfrac{1}{u^{2}}\sin^{2}\theta
\end{pmatrix}
\end{equation}
the required Christoffel symbols $\Gamma_{\nu\sigma}^{\mu}$ can be computed in a straightforward manner using the affine connection
\begin{equation}
\Gamma_{\nu\sigma}^{\mu}=\frac{1}{2} g^{\mu \rho}\left(\partial_{\nu} g_{\rho \sigma}+\partial_{\sigma} g_{\rho \nu}-\partial_{\rho} g_{\nu \sigma}\right).
\end{equation}
Thus, we obtain the non-zero $\Gamma_{\nu\sigma}^{\mu}$ as
\begin{align}
\Gamma_{t u}^{t}&=\Gamma_{u t}^{t}=\frac{1}{2} \frac{f'(u)}{f(u)} & &\nonumber
\\
\Gamma_{t t}^{u}&= \frac{1}{2} c^{2} u^{4}\,f(u)\,f'(u) & \Gamma_{u u}^{u}&=-\frac{1}{2} \frac{f'(u)}{f(u)}-\frac{2}{u},\nonumber
\\
\Gamma_{\theta \theta}^{u}&= u\,f(u) &\Gamma_{\phi \phi}^{u}&=u f(u) \sin^{2}\theta,\nonumber
\\
\Gamma_{u \theta}^{\theta}&=\Gamma_{\theta u}^{\theta}=-\frac{1}{u} & \Gamma_{\phi\phi}^{\theta}&=-\sin\theta\cos\theta \nonumber
\\
\Gamma_{u \phi}^{\phi}&=\Gamma_{\phi u}^{\phi}=-\frac{1}{u} & \Gamma_{\theta \phi}^{\phi}&=\Gamma_{\phi \theta}^{\phi}=\cot\theta.
\label{eq-chiristoffelsymbols}
\end{align}

\section{Conservation laws from the geodesic equations}
There are various ways to derive the conservation laws in relativistic mechanics. One elegant approach is to use the Killing vectors of the spacetime metric \cite[pp.~219\,ff.]{carroll:2019}. An equivalent method is to use the geodesic equations for the coordinates that do not appear explicitly in the metric components. In the Schwarzschild-(anti-)de Sitter metric given in \eqref{eq-schwartzschild1a}, these coordinates are $t$ and $\phi$. We begin by obtaining the first integrals of the geodesic equations, using $\tau$ as the affine parameter for these two coordinates (or $\lambda$ in the case of photons). By substituting the values of the Christoffel symbols given in Appendix C, it is straightforward to write equations \eqref{eq-geodesicseq1c} for $\mu=t$ and $\mu=\phi$ with $\lambda=\tau$ (or simply $\lambda$ for photons) as
\begin{align}
\frac{d}{d\tau}\left[f(u)\,\frac{dt}{d\tau}\right]&=0
\label{eq-geodesic-t-tau}
\\
\frac{d}{d\tau}\left[\frac{1}{u^{2}}\,\frac{d\phi}{d\tau}\right]&=0
\label{eq-geodesic-phi-tau}
\end{align}
which yield the conservation laws of energy, $E$, and angular momentum, $J$, respectively,
\begin{align}
f(u)\,\frac{dt}{d\tau}&=\frac{E}{c^{2}}
\label{eq-conservation-energy-tau}
\\
\frac{1}{u^{2}}\,\frac{d\phi}{d\tau}&=J
\label{eq-conservation-angularmomentum-tau}
\end{align}
For massive particles, $E$ and $J$ are conserved quantities per unit mass, whereas for photons they are simply constants of motion.

The conservation law for the angular momentum allows us to compute the pseudo-normalization condition given in equation \eqref{eq-norm-wmu} with $\lambda=\tau$ (or $\lambda$ for photons) and $\alpha=\phi$. This conservation law is the only relationship involving the proper time needed in the present derivation of Binet's equation.

Regarding again the non-affine formalism, equation \eqref{eq-geodesic-t-phi2} can also be integrated to obtain a conserved quantity related to $dt/d\phi$, which we will call $B$,
\begin{equation}
\frac{d}{d\phi}\left[u^{2} f(u)\,\frac{dt}{d\phi}\right]=0 \implies u^{2} f(u)\,\frac{dt}{d\phi}=B
\label{eq-conservation-t-phi}
\end{equation}
In effect, this equation allows the substitution of $dt/d\phi$ into \eqref{eq-geodesic-u-phi2}.

\section{Calculation of $\left(\mathrm{d}u/\mathrm{d}\phi\right)^{2}$ from the pseudo-normalization condition and the conservation laws}
The combination of the pseudo-normalization condition \eqref{eq-norm-wmu} with $\alpha=\phi$ and $\lambda=\tau$ (or $\lambda$ for photons) with equation \eqref{eq-conservation-t-phi} yields the square of the derivative of $u$ with respect to $\phi$
\begin{equation}
\left(\frac{d u}{d\phi}\right)^{2}=\begin{cases} c^{2}\,B^{2}-u^{2} f(u)-c^{2} u^{4} f(u)\left(\dfrac{d\tau}{d\phi}\right)^{2} & \mbox{(particle)} \\ \\ c^{2}\,B^{2}-u^{2} f(u) & \mbox{(photon)} \end{cases}
\label{eq-du-dphi-squared}
\end{equation}
\newline
The constant $B$ equals $E/(c^{2} J)$ as a consequence of \eqref{eq-conservation-energy-tau} and \eqref{eq-conservation-angularmomentum-tau}.

\section{Solution to the relativistic lightlike Binet's equation}

The convenience of using equation \eqref{eq-du-dphi-squared} instead of \eqref{eq-binet-relativistic1} —of which it is a first integral, even though we have not derived it as such— when finding the orbit equation lies in the known solvability of equations of the type
\begin{equation}
\left(y^{n}\,\frac{d y}{d x}\right)^{2}=P(y)
\label{eq-weierstrass}
\end{equation}
where $n=0,1$ and $P(y)$ is a polynomial in $y$ of degree no greater than $6$~\cite{hackmannthesis:2010}. The functional shape of both lightlike and timelike geodesics conforms to \eqref{eq-weierstrass}. The case of a null geodesic is simpler and corresponds to $n=0$ and a cubic polynomial $P(y)$,
\begin{equation}
\left(\frac{du}{d\phi}\right)^{2}=r_{S}u^{3}-u^{2}+C,
\label{eq:first-int}
\end{equation}
with $C=E^{2}/(c^{2}J^{2})+\Lambda/3$.

By making the change of variables 
\begin{equation}
u=\frac{4}{r_{S}}\,w+\frac{1}{3\,r_{S}}.
\label{eq:w-subst}
\end{equation}
one obtains
\begin{equation}
\left(\frac{dw}{d\phi}\right)^{2}=4\,w^{3}-g_{2}w-g_{3},
\label{eq:weierstrass}
\end{equation}
with the so-called \textit{Weierstrass invariants},
\begin{align}
g_{2}&=\frac{1}{12},\nonumber\\
g_{3}&=\frac{1}{216}-\frac{r_{S}^{2}}{16}\,C
=\frac{1}{216}-\frac{r_{S}^{2}}{16}\frac{E^{2}}{c^{2}J^{2}}-\frac{r_{S}^{2}}{48}\Lambda,
\label{eq:invariants}
\end{align}
for which the solution is the \textit{Weierstrass} $\wp$\textit{-function}:
\begin{equation}
w(\phi)=\wp\!\big(\phi-\phi_{\mathrm{in}};g_{2},g_{3}\big),
\end{equation}
so that the photon orbit is given by
\begin{equation}
r(\phi)=\frac{r_{S}}{4\,\wp\!\big(\phi-\phi_{\mathrm{in}};g_{2},g_{3}\big)+\tfrac{1}{3}\,}.
\label{eq:photon-solution}
\end{equation}
The phase constant $\phi_{\mathrm{in}}$ is fixed by the initial conditions. It is a non-trivial elliptic function and its full shape can be found in \cite{hackmannthesis:2010}.

\section*{Note}
This is the version of the article before peer review or editing, as submitted by an author to the \textit{European Journal of Physics} (IOP). This is a version of which all recognized typos have been corrected. IOP Publishing Ltd is not responsible for any errors or omissions in this version of the manuscript or any version derived from it. The Version of Record is available online at https://doi.org/10.1088/1361-6404/ae5af7

\bibliographystyle{unsrt}
\bibliography{references_binet}

\end{document}